# Neural Collaborative Filtering Classification Model to Obtain Prediction Reliabilities

Jesús Bobadilla*, Abraham Gutiérrez, Santiago Alonso, Ángel González-Prieto

Universidad Politécnica de Madrid, Dpto. Sistemas Informáticos, Madrid (Spain)

*Corresponding author: jesus.bobadilla@upm.es

**Abstract**

Neural collaborative filtering is the state of art field in the recommender systems area; it provides some models that obtain accurate predictions and recommendations. These models are regression-based, and they just return rating predictions. This paper proposes the use of a classification-based approach, returning both rating predictions and their reliabilities. The extra information (prediction reliabilities) can be used in a variety of relevant collaborative filtering areas such as detection of shilling attacks, recommendations explanation or navigational tools to show users and items dependences. Additionally, recommendation reliabilities can be gracefully provided to users: "probably you will like this film", "almost certainly you will like this song", etc. This paper provides the proposed neural architecture; it also tests that the quality of its recommendation results is as good as the state of art baselines. Remarkably, individual rating predictions are improved by using the proposed architecture compared to baselines. Experiments have been performed making use of four popular public datasets, showing generalizable quality results. Overall, the proposed architecture improves individual rating predictions quality, maintains recommendation results and opens the doors to a set of relevant collaborative filtering fields.

## 1 Introduction

Recommender Systems (RS) [1]-[2] are Artificial Intelligence systems designed to reduce the Internet information overload problem. RS can recommend items to users, avoiding large manual search to select appropriated products or services. Amazon, TripAdvisor, Netflix and Spotify are remarkable commercial firms that use RS. The selected filtering approaches are the RS core. The most relevant filtering methods are content-based [3] (books abstracts, products descriptions), demographic-based [4] (gender, age, zip), context-aware [5] (gps location), social [6] (followers, followed, tags), Collaborative Filtering (CF) [7]-[8] and hybrid [9] ensembles that joins two or more different filters. From the existing filtering approaches, the CF is the more relevant because it provides improved accuracy. From a machine learning point of view, historically CF has been addressed by using k-nearest neighbours (KNN), then Matrix Factorization (MF) [10] and currently Neural Collaborative filtering (NCF) [11]-[12]. Both MF and NCF models create an internal dense representation of each sparse user and item vector. In the first case we call to the representations: hidden factors, whereas in NCF they are embedding values. MF makes use of the same vector space to code users and items factors; NCF can be designed to make use of the same or different vector spaces. Finally, MF combines factors in a linear mode, whereas NCF combines embedding values in a non-linear mode, making it possible to catch the existing complex non-linear relations between users and items.

NCF has emerged in the RS area providing even better accuracy [13] than the traditional MF approaches [12]. DeepMF [14] is a general framework that implements MF by means of a neural model; it is a regression-based model implemented making use of two Multilayer Perceptrons (MLP) and a 'Dot' output layer. On several benchmark datasets, this model outperforms state of art machine learning models. The NCF [11] model extends the DeepMF [14] approach, replacing the output 'Dot' layer with an MLP and catching the complex non- linear relations between items and users. The NCF model proposed in [11] also introduces input embedding layers to make the model more scalable than the DeepMF [14] approach. Both the DeepMF and the NCF are regression-based architectures. The NCF area has been expanded to several fields beyond the RS domain; as an example, in [15] authors propose a new computational method NCFM (Neural network-based Collaborative Filtering Method) to predict miRNA- disease associations based on deep neural network. An automated and unsupervised method for the mitral valve segmentation using neural network collaborative filtering is explained in [16]. A framework to attack the QoS prediction in the IoT environment [17] combines NCF and fuzzy clustering; the NCF model is designed to leverage local and global features. A two sequential stages model (MF and neural) to improve fairness in RS [18] obtains fair recommendations without losing a significant proportion of accuracy. Additionally, a current RS for researchers and students: Deep Edu [19] makes use of NCF and it outperforms existing Educational services recommendation methods. The NCF input embedding layers have been refined in [20] by means of a user-item interaction graph. Among the remarkable current NCF approaches, we have selected a Joint Neural Collaborative Filtering (J_NCF) [21] that couples deep feature learning and deep interaction modelling with a rating matrix; the Contextual-boosted Deep Neural Collaborative filtering (CDNC) model [22] which simultaneously exploits both item introductions (textual features) and user ratings (collaborative features), making and ensembling of collaborative and content-based filtering; Knowledge graphs have been used to enhance NCF to alleviate the sparsity problem [23]; in [24] authors effectively combines user–item interaction information and auxiliary knowledge information for recommendation task; a NCF is proposed for user generated list recommendation, combining both item-level information and list-level information to improve performance. Finally, a neural embedding collaborative filtering (NECF) [25] is designed by using unsupervised auto-encoders, generating the embedding vectors from the user-item data, followed by a regression stage; as it can be seen, this is a DeepMF version where embeddings are replaced by auto-encoders.

An emerging beyond accuracy RS area is focused on the obtention of reliability values associated to the prediction ones; in this way, each prediction and recommendation will be represented by the pair <prediction, reliability>. The extra information (reliability) can be used to modulate recommendations: "you probably will like Avengers: Infinity War", "We really recommend you Avatar". Despite the reliabilities usefulness it has not been a main goal in the RS area: traditionally, CF has been excessively focused on accuracy. As a result, we use the number of votes as reliability measure: usually users take note of the number of people that has voted an item, and we prefer some four stars gadget voted by 1500 clients than other similar item voted five stars by 12 clients. Accurate CF methods to obtain prediction reliabilities can lead to nicer recommendation indications, such as the above examples. Additionally, the reliability information can be used for remarkable emerging areas such as detecting shilling attacks [26]: an unsupervised approach for detecting shilling attacks based on user rating behaviors [27] uses Dirichlet allocation model to extract latent topics of user preferences from user rating item sequences, whereas an MF approach to detect shilling attacks [28] tests the unusual reliability variations in the item predictions. Reliability values have also been used to make dynamic browsing of related users or items [29], to explain recommendations [30] and to filter to the most reliable recommendations [31]. A variety of methods and models have been proposed to get prediction and recommendation reliabilities; in a first stage trust-based [32] and similarity measures-based [33] methods were developed. Two remarkable machine learning approaches make use of MF ensembles, the first one [31] designs two sequential MF where the first one obtains prediction errors from known ratings, whereas the second MF make predictions from the previous errors, just getting the expected reliability values. The second MF approach [34] is the Bernoulli Matrix Factorization (BeMF), which is a matrix factorization model based on the Bernoulli distribution to exploit the binary nature of the designed classification model. Basically, BeMF runs a MF for each possible vote in the RS (e.g.: 1 to 5 stars), returning the probability (reliability) of each rating. The RS reliability field is growing fast due to the emerging reliability quality measures: a reliability quality prediction measure (RPI) and a reliability quality recommendation measure (RRI) are proposed in [35]. Both quality measures are based on the hypothesis that the more suitable a reliability measure is, the better accuracy results it will provide when applied. Current NCF is based on regression models [11, 12, 14, 19, 20, 21, 22], whereas our proposed model is a NCF architecture based on classification. This is an innovative approach whose RS accuracy must be compared to the conventional regression models. We have chosen the classification model because it presents a potential advantage: machine learning classification results provide probability distributions that could be used as reliability values in the CF context. Classification approaches naturally provide reliability values. A classification neural model for RS is proposed in [36], where the item relations patterns are learned in the network. It is designed to include an output layer containing as many neurons as items in the dataset, what usually is an affordable approach, although it could be considered as not scalable in specific scenarios; nevertheless, it provides prediction reliabilities. In [37] two sequential models have been designed to exploit the potentiality of the reliability information: the first model uses MF to obtain reliabilities, whereas the second one is a neural model that improves recommendation accuracy by incorporating in its input layer the previous MF reliability values.

The proposed classification architecture in this paper borrows the NCF classification concept from [36] and the NCF design from [11], trying to catch the individual strengths of both approaches. It also incorporates two innovative contributions to improve scalability through the use of input embeddings and to improve accuracy by replacing the usual regression 'Dot' layer with a 'Concatenate' layer that preserves the abstract information from hidden layers. By running the proposed model, we obtain pairs <prediction value, prediction reliability> and we make use of the reliability 'extra' information to get the most promising recommendations. Since the proposed classification architecture returns prediction reliabilities, it opens the door to remarkable state of art research fields such as mentioned above. The key question now is: will the classification approach return less accurate recommendations than the state of art regression approaches? If the answer is affirmative our classification architecture loses its purpose, whereas if it returns similar or higher accuracy than regression NCF it is valuable to use it and to take advantage of the reliability additional information. Thus, our hypothesis is that the proposed classification architecture provides similar or higher accuracy than the current NCF regression models, making it possible to take advantage of the extra reliability information it returns. The rest of the paper has been structured as follows: in Section II the proposed model is explained, and the experiments design is defined. Section III shows the experiments' results and their discussions. Finally, Section IV contains the main conclusions of the paper and the future works.

## 2 Materials and Methods

This section contains two different subsections: the first one defines and explains the design of the proposed architecture and its relationship with the state of art models. The second subsection focuses on the experiments design and their implementation, defining the selected datasets, the chosen baselines, the stablished parameter values and the tested quality measures.

### 2.1 The Proposed Classification Based Neuronal Architecture

The proposed classification architecture has been designed following the DeepMF and the NCF state of art evolution, taking those elements that are more advantageous to the classification-based approach and removing the ones that are not appropriate. From the DeepMF [14] we have borrowed the two designed MLP branches to separately process item and user's information. Fig. 1 shows the DeepMF architecture with its two separate N layers MLP neural networks. However, this DeepMF design has a remarkable drawback: it has not an appropriate scalability.

As it can be seen in Fig. 1, both MLP branches are fed by using very large vectors as input: each user vector of ratings and each item vector of ratings. Please note that each item vector can contain millions of ratings, since some RSs contain explicit or implicit ratings coming from millions of users.

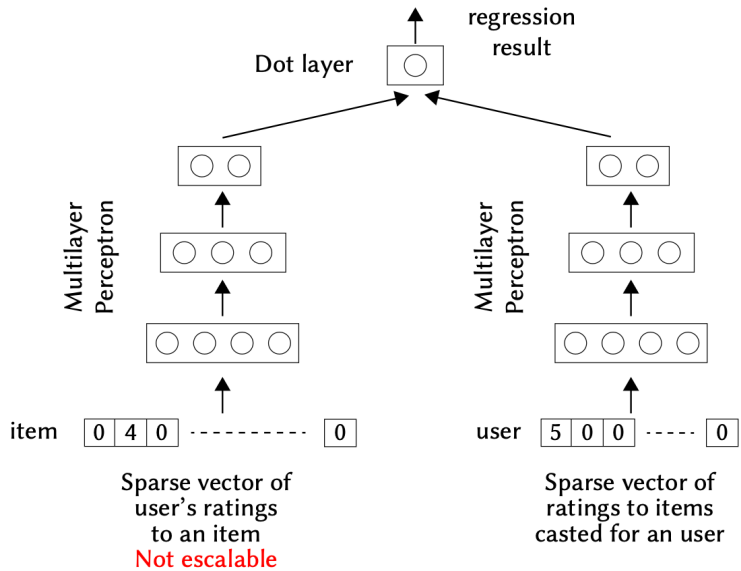

Figure 1: DeepMF architecture [14].

Our classification-based solution makes use of an optimized approach to feed the neural network: the use of embedding layers. This architectural solution has been used in the NCF [11] design as shown in Fig. 2: now the input layer is coded by using vectors of bits that are processed in the embedding layers, providing both the user and the item latent vectors. Additionally, and MLP neural network is used to process the latent vectors instead of the "Dot" layer used in the DeepMF approach; this MLP makes possible to find the complex non- linear dependences existing among the embedding values.

Our proposed architecture is shown in Fig. 3; its inputs are just two numbers (for each existing rating): the item code and the user code. These two numbers feed two separate embedding layers that code each number as a vector of F values. The embedding process assigns similar embedding layer vector values to similar users (or similar items), making much easier the subsequent MLP tasks. The embedding layer operations make use of a lookup table where the keys are the users (or items) numbers, and the values are the dense vectors of size F. In this way we do not longer need to maintain explicit one-hot encoder sparse vectors as the ones shown in the NCF architecture (Fig. 2). Embedding layers are mainly used in natural language processing scenarios due to the huge sparsity of the words in sentence representations. RS datasets also present high sparsity levels, since users only cast ratings from a very reduced proportion of the available items. The embedding layer can be instanced by providing the maximum number of existing elements (users or items in the CF context) and the size of the compressed information (usually from 5 to 20 values in the CF context). The proposed architecture embedding layer for items receives as input each item number from each existing rating. Likewise, the user's embedding layer receives as input each user number from each existing rating. Each input number of the sequence is used as index to access a lookup table (embedding weight matrix) containing vectors for each user (or item). The lookup tables efficiently implement the embedding layers. Both the user and the item embedding layers compress and code each item and user ID (number). Once the embedding training process is finished, in the same way in which NLP related words have close embedding representations, CF related users and items have close embedding representations. Since our proposed architecture is classification based, the output layer has V neurons (Fig. 3), in contrast to the regression NCF version (Fig. 2) where one single neuron makes the regression. The V value depends on the number of different available implicit ratings or explicit votes in the RS (e.g.: 1, 2, 3, 4 or 5 stars). The output of the multilayer perceptron is positional (one neuron for each rating value, in this case 5-stars). Each output neuron provides a reliability measure. Whereas the NCF classification loss function is binary, in our classification architecture we use a categorical cross entropy loss; thus, our classification output layer returns V probabilities: $v_i \in \mathcal{R}$, $i \in \{1, 2, ..., V\}$, $\sum_i v_i = 1$. Please note that the $v_i$ value's argmax function provides us with the discrete prediction, whereas the corresponding $v_i$ value can be seen as the prediction reliability. This is a very different scenario to the regression based NCF, where prediction results are continuous prediction values. The classification NCF results are richer than the regression ones and it opens the door to use this additional information to deal with diverse RS goals such as providing prediction reliabilities, improving recommendation quality, making items or users relations graphs, or explaining recommendations.

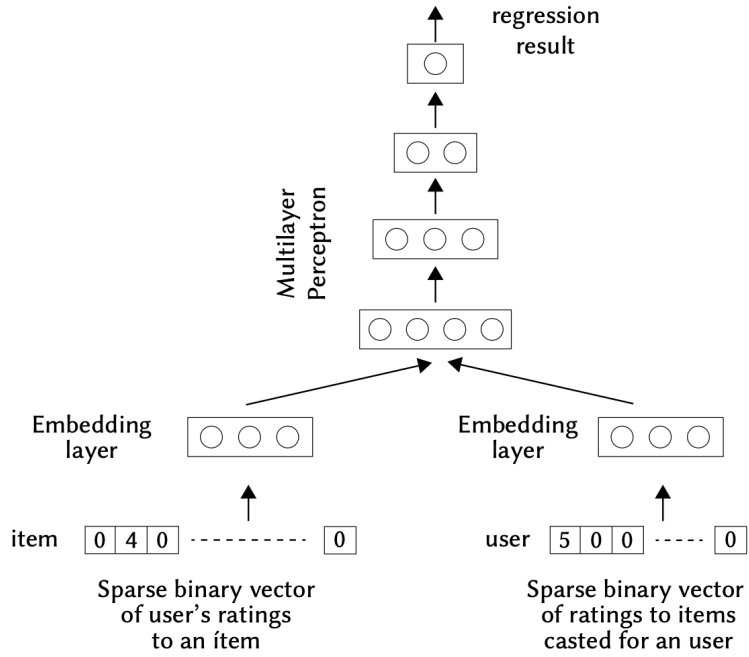

Figure 2: NCF architecture [11].

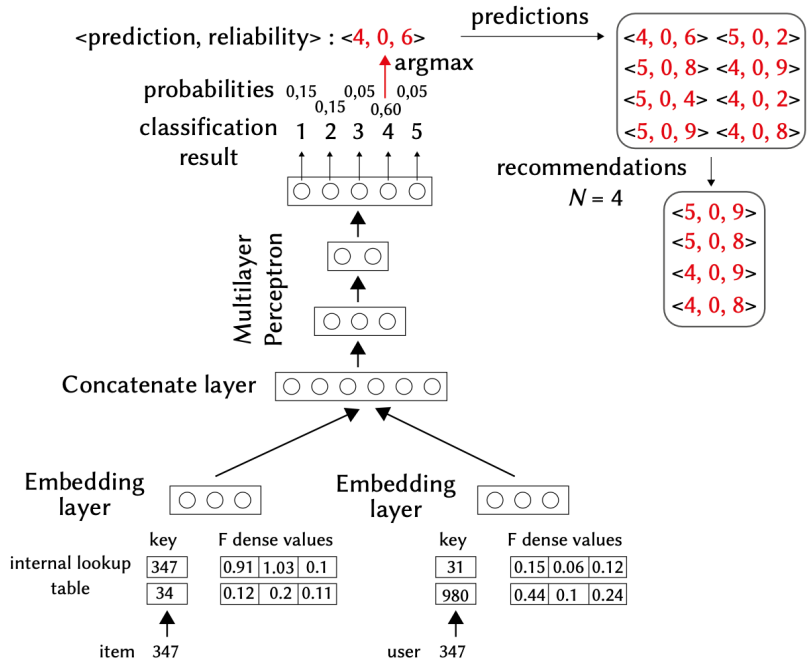

Figure 3: Proposed architecture and recommendation method.

In short, we encourage the use of an NCF classification-based architecture whose samples are simple item and user numbers, and whose output labels are ratings. This design avoids large input vectors, and it returns <rating, reliability> prediction pairs. Hereafter we provide an example to show the flexibility of the classification based NCF: we will choose the most promising recommendations from their rich results, compared to the restricted operation evolved in the regression approach. From the following set of pairs <prediction, reliability>: <5, 0.3>, <5, 0.2>, <5, 1>, <5, 0.9>, <4, 0.8>, <4, 0.4>, <4, 0.7>, <3, 0.7>, the following ordered list of recommendations could be obtained: <5, 1>, <5, 0.9>, <4, 0.8>, <4, 0.7>. We have just filtered to the highest predictions (4 & 5) and the highest reliabilities (reliability >= 0.5), and then we have ordered the resulting pairs attending to their reliability. As it can be seen, risky recommendations have been avoided (<5, 0.2>, <5, 0.3>, <4, 0.4>) in a process that is not available using the NCF regression results.

The Algorithm 1 Keras code section shows a specific implementation of the proposed NCF classification architecture. In this case, the dataset contains 5 categories (1 to 5 stars) and the chosen embedding size is 10 (line 1). Lines 2 to 4 input each movie number code, embedding them and prepare their flatten representation; lines 5 to 7 make the same process for each user number code. Both the movie and user flatten representations are concatenated in the 'Concatenate' layer in line 8. The existing complex non-linear relations between users and items are learnt in the hidden layers coded from line 9 to line 12. Line 13 creates the five categories classification output layer, and it assigns the softmax activation function to the neurons. The whole model is created in line 14, defining the users/items input tensor and the output layer. Finally, the compile and fit methods are set in lines 15 and 16.

Algorithm 1. Classification NCF design

```
1. embed_size = 10; num_categ = 5

2. movie_input = Input(shape=[1])
3. movie_embedding = Embedding(num_movies + 1, embed_size)(movie_input)
4. movie_flatten = Flatten()(movie_embedding)

5. user_input = Input(shape=[1])
6. user_embedding = Embedding(num_users + 1, embed_size)(user_input)
7. user_flatten = Flatten()(user_embedding)

8. concat = Concatenate(axis=1)([movie_flatten, user_flatten])
9. mlp_1 = Dense(80, activation='relu')(concat)
10.mlp_2 = Dropout(0.4)(mlp_1)
11.mlp_3 = Dense(25, activation='relu')(mlp_2)
12.mlp_4 = Dropout(0.4)(mlp_3)
13.output = Dense(num_categ, activation='softmax')(mlp_4)

14.model_classification = Model([user_input, movie_input], output)
15.model_classification.compile( optimizer='adam', metrics=['mae'],
        loss='categorical_crossentropy')
16.history = model_classification.fit( [train[:,USER], train[:,ITEM]],
        to_categorical(train[:,RATING]),
        validation_data=([test[:,USER],test[:,ITEM]],
        to_categorical(test[:,RATING])), epochs=EPOCHS, verbose=1)
```

## 2.2 Experiments Design

This paper's hypothesis claims that classification based NCF provides similar or better accuracy than the regression based NCF model. Additionally, classification based NCF allows to tackle a variety of RS goals in a simpler way than the regression approach. To test the comparative accuracy of both NCF approaches we have designed a set of experiments involving prediction and recommendation results, processing a variety of baselines and using several public CF datasets. The selected RS datasets are: Movielens 100K [38], Movielens 1M [38], MyAnimeList* (a subset of the original dataset) [39] and Netflix* [40] (a subset of the original dataset). Table I shows these datasets main parameter values.

| Dataset | #users | #items | #ratings | scores | sparsity |
|---|---|---|---|---|---|
| **Movielens 100K** | 943 | 1682 | 99,831 | 1 to 5 | 93,71 |
| **MovieLens 1M** | 6,040 | 3,706 | 911,031 | 1 to 5 | 95,94 |
| **MyAnimeList*** | 19,179 | 2,692 | 548.967 | 1 to 10 | 98,94 |
| **Netflix*** | 23,012 | 1,750 | 535,421 | 1 to 5 | 98,68 |

Table I. Main Parameters of the Datasets Used in the Experiments

The paper's baselines are: DeepMF [14], NCF regression [11], NCF classification, and binary NCF classification. NCF classification corresponds to the architecture in Fig. 3, whereas binary NCF classification is the NCF classification version where ratings have been converted to "relevant" or "not relevant" (e.g.: relevant ⇔ rating >= 4, not relevant ó rating < 4), and labels have also been converted to the exposed 'relevant' and 'not relevant' discrete classification. The difference between NCF classification and the proposed method is that reliability results are not used in the baseline. Table II abstracts both the proposed and the baselines architectures. Both the DeepMF and the regression architectures make use of a 'Dot' layer to join their embeddings, whereas a 'Concatenate' layer has been used in both the proposed and the baseline classification architectures.

| Architecture | type | merge layer | Reliability inf. |
|---|---|---|---|
| **Proposed** | classification | concatenate | yes |
| **DeepMF** | regression | dot | no |
| **Regression** | regression | dot | no |
| **Classification** | classification | concatenate | no |
| **Binary classification** | classification | concatenate | no |

Table II. Proposed and Baseline Architectures

The main experiments make use of the precision and recall recommendation measures, tested by using a range from 2 to 10 number of recommendations (N). A secondary set of experiments tests the prediction quality on each of the dataset ratings (1 to 5 stars), using 2, 6 or 10 number of recommendations (7, 8 and 9 for the MyAnimeList dataset). Finally, an experiment tests the precision versus coverage obtained by filtering to predictions higher than a beta threshold. In the CF context, the relevancy threshold parameter is used to classify a rating or a prediction in the categorical set {relevant, not_relevant}. Experiments on datasets where votes range from 1 to 5 (Movielens, Netflix, etc.) usually set the relevancy threshold in the value 4; that is: ratings 4 and 5 are considered relevant, and predictions greater than or equal to 4 are considered as candidates to be recommended. The relevancy threshold is also important in the cross-validation testing process, since recommendations made to items voted under 4 are considered as errors, whereas recommendations made to items voted 4 or above are considered as successes. In this example, setting the relevancy threshold to 5 makes it more difficult to get high success (precision, recall, etc.) levels. Table III shows the experiments and their main parameter values.

| Experiment | # recomm. (N) | Relevancy | Prediction value |
|---|---|---|---|
| **Precision / recall** | {2,4,6, …, 10} | {3, 4, 5} Movielens and Netflix {7, 8, 9} MyAnimeList | {0} |
| **Quality predicting each rating {1, 2, …, 5}** | {2, 6, 10} Movielens and Netflix {7, 8, 9} MyAnimeList | {4} | {0} |
| **Precision vs. coverage** | {10} | {3, 4, 5} | {4, 4.2, 4.4, 4.6, 4.8} |

Table III. Experiments and Their Main Parameter Values

## 3 Results

Our first experiment compares the quality of recommendations obtained by using the proposed method and the baselines. The chosen CF quality measures have been the precision and recall ones. The main parameters of this experiment are abstracted in the Table III first row. Fig. 4 shows the results obtained on both the Movielens 100K (top graphs) and Movielens 1M (bottom graphs). The main conclusion is that, as expected, the proposed classification based NCF method reaches adequate recommendation quality results, compared to state of art NCF baselines. Additionally, we can see that both the recall and the precision evolutions are the expected ones: precision decreases as the number of recommendations (N) increases, whereas recall increases as the N value increases. The y axis scale shows better results in the graphs corresponding to the 1M Movielens version compared to the 100K one. Fig. 4 results are particularly interesting when the recommendations difficulty is increased: when the relevancy threshold is set to its limit (value 5); in this scenario we find a Movielens 100K optimum number of recommendations: 7 (graph on the top-right of Fig. 4), an also a remarkable behaviour of the proposed method when applied to Movielens 1M (graph on the bottom-right of Fig. 4).

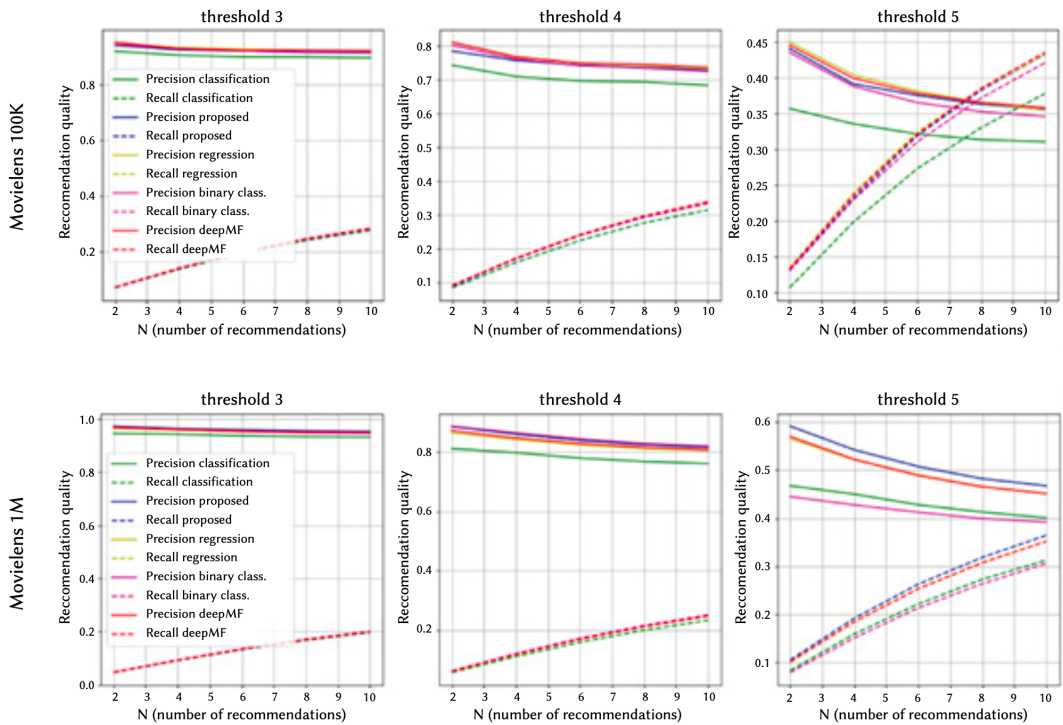

Figure 4: Precision and recall quality results obtained by using the Movielens 100K (top graphs) and the Movielens 1M (bottom graphs). Three relevancy thresholds have been tested: 3 (left graphs), 4 (center graphs), 5 (right graphs). The proposed architecture has been compared to the baselines: deepMF [14], NCF regression [11], regular NCF classification and binary NCF classification. Precision and recall "proposed" values are the results of the proposed method in the paper (Table II).

The same kind of experiments have been performed by using the MyAnimeList* and Netflix* datasets (Table I). Results confirm the conclusions obtained in both Movielens datasets, particularly it is confirmed that the proposed NCF classification approach reaches adequate recommendation results. The top graphs of Fig. 5 show the MyAnimeList* results; this dataset has a range of votes from 1 to 10, instead the usual 1 to 5. This circumstance reduces the quality of the binary classification baseline, since the binary differentiation between relevant and not relevant ratings becomes less precise.

Fig. 5 top-center and top-right graphs shows the mentioned quality drop. It is also remarkable that in both Fig. 4 and Fig. 5 we can observe a better behavior of the NCF classification proposed approach than the NCF classification regular method; this means that an adequate use of the reliability information leads to the expected quality improvements.

The explained experiments are based on the usual approach to test recommendation quality: proportion of the relevant recommended items versus the total number of recommendations (precision) or versus the total number of relevant items (recall). In both cases, we are considering a hit if the recommended item exceeds a relevancy threshold. In some RS scenarios we need a more fine-grained predictions and recommendations. This is the case when we want to be sure we will not like an item, or we will only like it if it has been recommended with a particular value, e.g.: I am interested in a set of songs, but I want to discard everyone who has obtained 1 or 2 stars. I want to watch some films unknown to me, then I will surf by those the RS recommendation is three stars. To test the proposed and the baselines NCF approaches in this scenario, the experiments shown in the second row of Table III have been run. Basically, we obtain the quality of each architecture to appropriately recommend each existing rating in the CF dataset. The binary NCF classification has not been included since it is not designed to predict individual ratings.

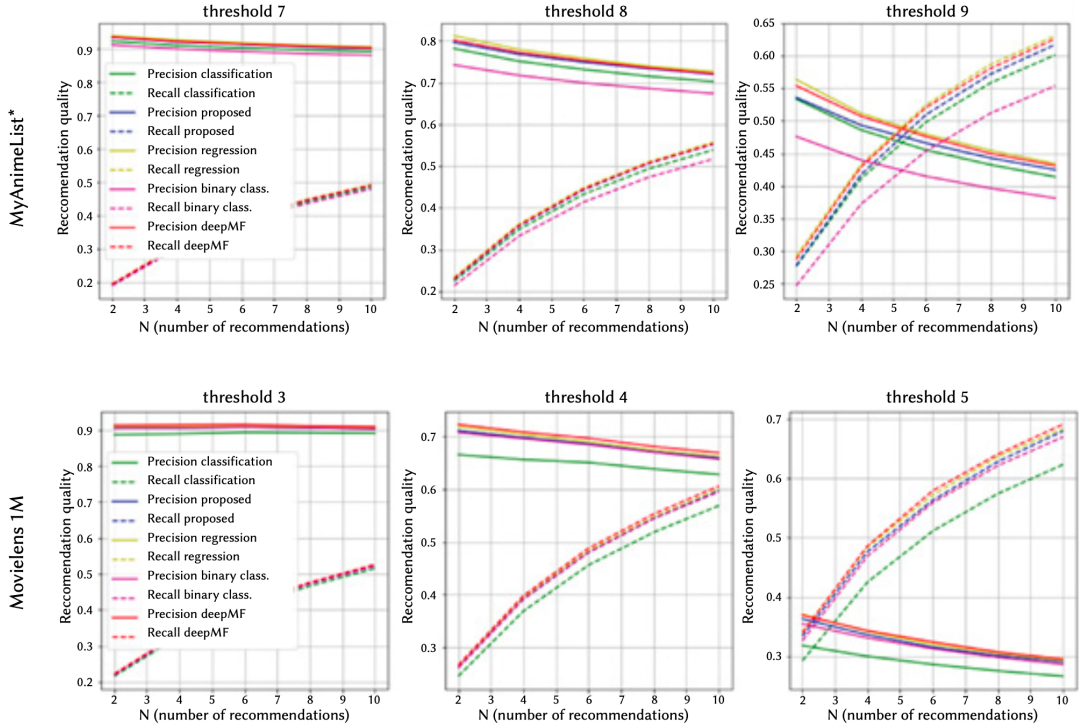

Fig. 5 Precision and recall quality results obtained by using the MyAnimeList* (top graphs) and the Netfix* (bottom graphs). Three relevancy thresholds have been tested: 7 or 3 (left graphs), 8 or 4 (center graphs), 9 or 5 (right graphs). The proposed architecture has been compared to the baselines: DeepMF [14], NCF regression [11], regular NCF classification and binary NCF classification. Precision and recall "proposed" values are the results of the proposed method in the paper (Table II).

Fig. 6 shows the obtained results on the covered datasets. Beyond details, it can be seen the superiority of the proposed NCF method in all the contemplated scenarios. This is the expected result, since DeepMF and NCF regression just provide prediction values, whereas the proposed NCF classification architecture returns specific and complete classification information shaped like <rating, reliability> pairs. The remarkable differences in the quality results of the different ratings (x-axis) when predicted by the NCF classification or regression methods is due to the CF datasets are usually biased to high rating values. As an example: both Movielens datasets show a much bigger difference between NCF classification and regression methods for the rating 1 and the rating 2 predictions; users tend to cast high votes (4 and 5), and then the number of low votes (1 and 2) are usually scarce in CF datasets. In this scenario, the NCF classification method particularly shows its superiority over the NCF regression one.

Finally, a set of experiments have been conducted (third row in Table III) to compare precision and coverage in the NCF scenario. It is important to realize that several parameters determine the recommendation coverage in a cross-validation scenario: a) the dataset distribution of ratings, particularly the rating matrix sparsity, b) the requested number of recommendations (N), c) the required threshold (4 stars, 5 stars); e.g.: it can be difficult to find users to recommend N=10 items with a 5 threshold in a cross-validation testing scenario. Usually, recommendation quality and coverage are inversely related. To conduct the experiments, we introduce a beta threshold used to select predictions that are greater than beta, e.g.: centered graph in Fig. 7 shows recommendation results where the relevancy threshold is 4 stars (top label); looking at the x-axis in its 4.4 beta threshold we can know the obtained coverage and recommendation precision selecting predictions equal to or higher than 4.4. As it can be seen, by increasing the beta value we also increase precision, but at the cost of a sharp decrease of the coverage. In the current experiment, the key question is to compare the precision versus coverage equilibrium in both the proposed method and the baselines. We can observe that the proposed method gets and intermediate position between DeepMF and NCF regression, and we can conclude that the baselines and the proposed method similarly performs in this particular issue: thus, the proposed classification architecture use does not worsen the RS coverage.

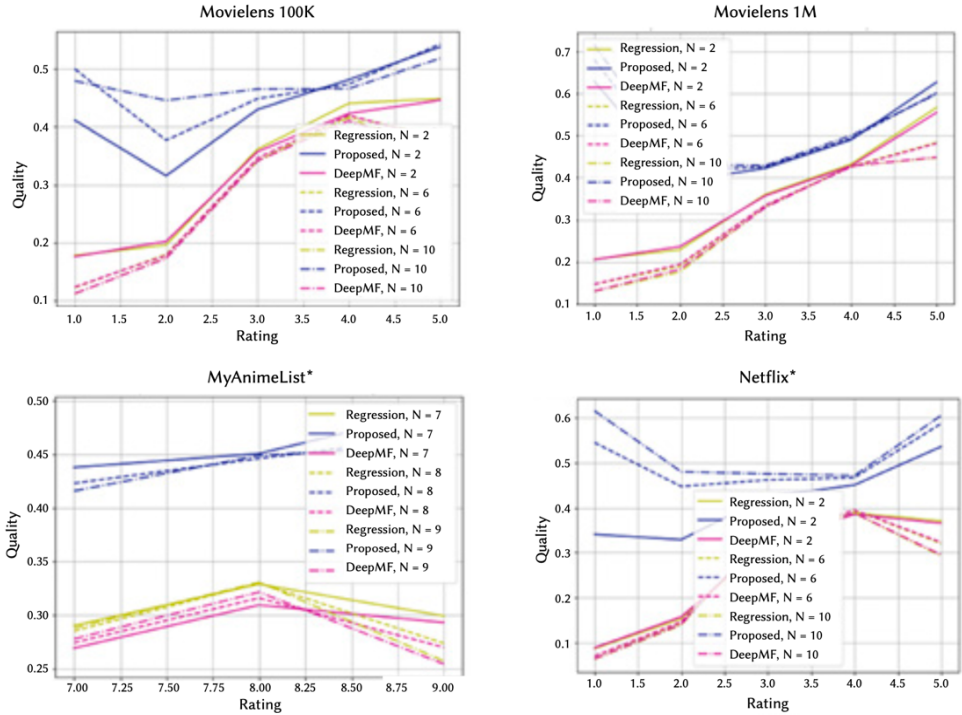

Fig. 6. Obtained precision quality recommending items for each considered rating, e.g.: proportion of hits recommending items that the user has voted with 2 stars. Top-left graph: Movielens 100K, top-right graph: Movielens 1M, bottom-left graph: MyAnimeList*, bottom-right graph: Netfix*. x-axis: ratings (7, 8 and 9 for MyAnimeList*; 1, 2, 3, 4 and 5 in Netfix* and Movielens). The proposed architecture has been compared to the baselines: DeepMF [14], NCF regression [11]. "Proposed" values are the results of the proposed method in the paper (Table II).

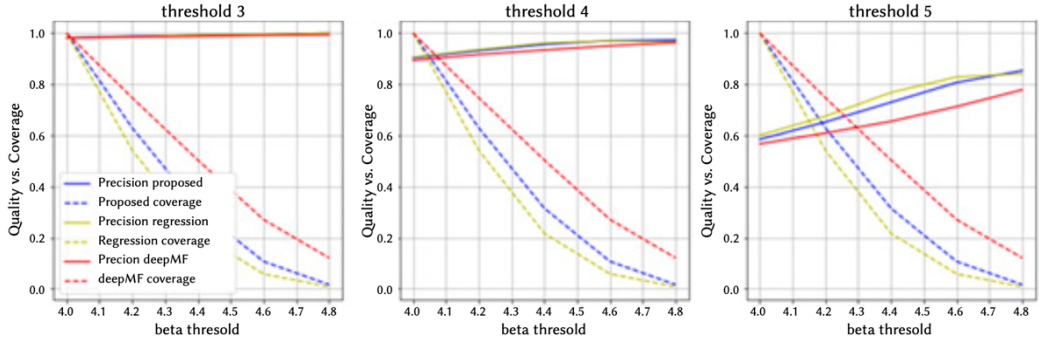

Fig. 7. Precision versus coverage results. Dataset: Movielens 1M; number of recommendations (N): 10; relevancy threshold: 3 stars (left graph), 4 stars (center), 5 stars (right graph); x-axis: beta threshold (selects predictions equal or higher than beta). Precision and coverage "proposed" values are the results of the proposed method in the paper (Table II).

## 4 Conclusion

Current collaborative filtering deep learning architectures are focused on the regression approach; they provide accurate predictions and recommendations compared to the state of art, but they do not return any reliability value of such predictions and recommendations. Conversely, classification based deep learning architectures are able to provide both the value and the reliability of each prediction or recommendation. By combining the prediction value and the reliability information it is possible to afford several remarkable tasks, such as obtaining more reliable recommendations, making some reliability- based explanations of the recommendations to users, showing navigable trees that relate users or items, implementing methods to reduce the shilling attacks consequences, etc.

Despite the mentioned advantages of the classification-based neural collaborative filtering approach, the proposed approach cannot be embraced without testing its quality performance: similar or better results must be obtained by applying the proposed architecture compared to the state of art regression baselines. Experiments in this paper show that the proposed classification architecture obtains similar recommendation accuracy results than the regression architectures do; precision and recall measures provide comparable quality results in a complete set of experiments where different thresholds and diverse number of recommendations are chosen.

Results show a consistent pattern when experiments have been run on four public datasets. On the other hand, the proposed architecture shows improved prediction results: it is able to accurately predict individual ratings, outperforming prediction quality compared with the state of art regression approaches. Finally, the precision versus coverage balance stays similar in both the proposed and the baselines neural architectures.

In short, the proposed classification-based architecture can replace the state of art neural collaborative filtering approaches: its use does not worsen the recommendation quality, it improves the prediction of individual ratings, and it opens the door to a set of relevant collaborative filtering areas. Remarkable future works from this paper are: to make use of reliabilities to detect shilling attacks, to provide reliability values in the users' recommendations, and to filter non reliable recommendations.

# Acknowledgements

This work was partially supported by *Ministerio de Ciencia e Innovación* of Spain under the project PID2019-106493RB-I00 (DL-CEMG) and the *Comunidad de Madrid* under *Convenio Plurianual* with the Universidad Politécnica de Madrid in the actuation line of *Programa de Excelencia para el Profesorado Universitario*.

# Conflict of interest

The authors declare that they have no conflict of interest.

# References

[1] K. Madadipouya, S. Chelliah, "A Literature Review on Recommender Systems Algorithms, Techniques and Evaluations", Brain: Broad Research in Artificial Intelligence and Neuroscience, vol. 8, no. 2, 2017, pp. 109-124.

[2] S.S. Sohail, J. Siddiqui, R. Ali, "Classifications of Recommender Systems: A review", Journal of Engineering Science and Technology Review, vol. 10, no. 4, 2017, pp. 132-153.

[3] H. Zamani, A. Shakery, "A language model-based framework for multi- publisher content-based recommender systems", Information Retrieval Journal, vol. 21, no. 5, 2018, pp. 369-409.

[4] M.Y.H. Al-Shamri, "User profiling approaches for demographic recommender systems", Knowledge-Based Systems, vol. 100, 2016, pp. 175-187.

[5] N.M. Villegas, C. Sánchez, J. Díaz-Cely, G. Tamura, "Characterizing context-aware recommender systems: A systematic literature review", Knowledge-Based Systems, vol. 140, 2018, pp. 173-200.

[6] Rezvanian, B. Moradabadi, M. Ghavipour, M.M. Daliri Khomami, M.R. Meybodi, "Social recommender systems", Studies in Computational Intelligence, vol. 820, 2019, pp. 281-313.

[7] Hernando, J. Bobadilla, F. Ortega, A. Gutiérrez, "A probabilistic model for recommending to new cold-start non-registered users", Information Sciences, vol. 376, 2017, pp. 216-232.

[8] J. Bobadilla, A. Gutiérrez, S. Alonso, R. Hurtado, "A Collaborative Filtering Probabilistic Approach for Recommendation to Large Homogeneous and Automatically Detected Groups", International Journal of Interactive Multimedia and Artificial Intelligence, 2020, doi: 10.9781/ijimai.2020.03.002.

[9] V. Yu. Ignat'ev, D. V. Lemtyuzhnikova, D. I. Rul', I. L. Ryabov, "Constructing a Hybrid Recommender System", Journal of Computer and Systems Sciences International, vol. 57, no. 6, 2018, pp. 921-926.

[10] H. Li, Y. Liu, Y. Qian, N. Mamoulis, W. Tu, Wenting ; D. Cheung, "HHMF: hidden hierarchical matrix factorization for recommender systems", Data Mining and Knowledge Discovery, vol. 33, no. 6, 2019, pp. 1548-1582.

[11] H. Xiangnan, L. Lizi, Z. Hanwang, "Neural Collaborative Filtering", in International World Wide Web Conference Committee (IW3C2), Perth, Australia, 2017, doi: 10.1145/3038912.3052569

[12] D. Bokde, S. Girase, D. Mukhopadhyay, "Matrix Factorization Model in Collaborative Filtering Algorithms: A Survey", Procedia Computer Science, vol. 49, 2015, pp. 136-146, doi: 10.1016/j.procs.2015.04.237.

[13] S. Rendle, W. Krichene, L. Zhang, J.R. Anderson, "Neural Collaborative Filtering vs. Matrix Factorization", in RecSys '20: Fourteenth ACM Conference on Recommender Systems, Brasil, 2020, pp. 240–248, doi: 10.1145/3383313.3412488.

[14] H.J. Xue, Xi. Dai, J. Zhang, S. Huang, J. Chen, "Deep Matrix Factorization Models for Recommender Systems", in Proceedings of the Twenty-Sixth International Joint Conference on Artificial Intelligence, Melbourne, Australia, 2017, pp. 3203-3209, doi: 10.24963/ijcai.2017/447

[15] Y. Liu, S.L. Wang, J.F. Zhang, W. Zhang, W. Li, "A neural collaborative filtering method for identifying miRNA-disease associations", Neurocomputing, vol. 422, 2021, pp. 176-185.

[16] L. Corinzia, F. Laumer, A. Candreva, M. Taramasso, F. Maisano, J.M. Buhmann, "Neural collaborative filtering for unsupervised mitral valve segmentation in echocardiography", Artificial intelligence in medicine, vol. 110, 2020, pp. 101975-101975.

[17] H. Gao, Y. Xu, Y. Yin, W. Zhang, R. Li, X. Wang, "Context-Aware QoS Prediction with Neural Collaborative Filtering for Internet-of-Things Services", IEEE internet of things journal, vol. 7, no. 5, 2020, pp. 4532-4542, doi: 10.110'9/JIOT.2019.2956827.
[18] J. Bobadilla, R. Lara-Cabrera, A. González-Prieto, F. Ortega, "DeepFair: Deep Learning for Improving Fairness in Recommender Systems", International Journal Of Interactive Multimedia And Artificial Intelligence, 2020, doi: 10.9781/ijimai.2020.11.001.
[19] F. Ullah, B. Zhang, R.U. Khan, T.S. Chung, M. Attique, K. Khan, S. Khediri, S. Jan, "Deep Edu: A Deep Neural Collaborative Filtering for Educational Services Recommendation", IEEE access, vol. 8, 2020, pp. 110915-110928.
[20] Y. Guo, Z. Yan, "Recommended System: Attentive Neural Collaborative Filtering", IEEE access, vol. 8, 2020, pp. 125953-125960.
[21] W. Chen, F. Cai, H. Chen, M. Rijke, "Joint Neural Collaborative Filtering for Recommender Systems", ACM transactions on information systems, vol. 37, no. 4, 2019, pp. 1-30.
[22] S. Yu, M. Yang, Min, Q. Qu, Y. Shen, "Contextual-boosted deep neural collaborative filtering model for interpretable recommendation", Expert systems with applications, vol. 136, 2019, pp. 365-375.
[23] L. Sang, M. Xu, S. Qian, X. Wu, "Knowledge graph enhanced neural collaborative recommendation", Expert systems with applications, vol. 164, 2021, pp. 113992, doi: 10.1016/j.eswa.2020.113992.
[24] C. Yang, L. Miao, B. Jiang, D. Li, D. Cao, "Gated and attentive neural collaborative filtering for user generated list recommendation", Knowledge-based systems, vol. 187, 2020, pp. 104839.
[25] T. Huang, D. Zhang, L. Bi, "Neural embedding collaborative filtering for recommender systems", Neural computing & applications, vol. 32, no. 22, 2020, pp. 17043-17057.
[26] M. Si, Q. Li, "Shilling attacks against collaborative recommender systems: a review", The Artificial intelligence review, vol. 53, no. 1, 2018, pp. 291-319.
[27] F. Zhang, Z. Ling, S. Wang, "Unsupervised approach for detecting shilling attacks in collaborative recommender systems based on user rating behaviours", IET information security, vol. 13, no. 3, 2019, pp. 174-187.
[28] S. Alonso, J. Bobadilla, F. Ortega, R. Moya, "Robust Model-Based Reliability Approach to Tackle Shilling Attacks in Collaborative Filtering Recommender Systems", IEEE access, vol. 7, 2019, pp. 41782-41798.
[29] Hernando, J. Bobadilla, F. Ortega, A. Gutiérrez, "Method to interactively visualize and navigate related information", Expert Systems with Applications, vol. 111, 2018, pp. 61-75.
[30] Hernando, R. Moya, F. Ortega, J. Bobadilla, "Hierarchical graph maps for visualization of collaborative recommender systems", Journal of Information Science, vol. 40, no. 1, 2014, pp. 97-106.
[31] Zhu, F. Ortega, J. Bobadilla, A. Gutiérrez, "Assigning reliability values to recommendations using matrix factorization", Journal of computational science, vol. 26, 2018, pp. 165-177.
[32] S. Ahmadian, P. Moradi, F. Akhlaghian, Fardin, "An improved model of trust-aware recommender systems using reliability measurements", in 6th Conference on Information and Knowledge Technology (IKT), Shahrud, Iran, 2014, pp. 98-103.
[33] Hernando, J. Bobadilla, F. Ortega, J. Tejedor, "Incorporating reliability measurements into the predictions of a recommender system", Information Sciences, vol. 218, 2013, pp. 1-16.
[34] F. Ortega, R. Lara-Cabrera, A. González-Prieto, J. Bobadilla, "Providing reliability in recommender systems through Bernoulli Matrix Factorization", Information sciences, vol. 553, 2021, pp. 110-128.
[35] J. Bobadilla, A. Gutiérrez, F. Ortega, B. Zhu, "Reliability quality measures for recommender systems", Information Sciences, Vol. 442-443, 2018, pp. 145-157.
[36] J. Bobadilla, F. Ortega, A. Gutierrez, S. Alonso, "Classification-based Deep Neural Network Architecture for Collaborative Filtering Recommender Systems", International Journal of Interactive Multimedia and Artificial Intelligence, vol. 6, no. 1, 2020, pp. 68-77.
[37] J. Bobadilla, S. Alonso, A. Hernando, "Deep learning architecture for collaborative filtering recommender systems", Applied Sciences, vol. 10, no. 7, 2020, pp. 2441.
[38] F.M. Harper, J.A. Konstan, "The movielens datasets: History and context",ACM Transactions on Interactive Intelligent Systems, vol. 5, no. 4, 2015, pp. 1–19.
[39] https://www.kaggle.com/azathoth42/myanimelist
[40] F. Ortega, B. Zhu, J. Bobadilla, A. Hernando, "CF4J: Collaborative filtering for Java", Knowledge-Based Systems, vol. 152, 2018, pp. 94–99.